\documentclass{article}

\usepackage{microtype}
\usepackage{amsmath}
\usepackage{graphicx}
\usepackage{subfigure}
\usepackage{booktabs}
\usepackage{hyperref}

\usepackage[accepted]{icml2022}
\icmltitlerunning{Adversarially-trained artificial neurons are more robust than biological neurons}

\begin{document}
\twocolumn[
\icmltitle{Adversarially trained neural representations may already be as robust as corresponding biological neural representations}

\begin{icmlauthorlist}
\icmlauthor{Chong Guo}{mc1}
\icmlauthor{Michael J. Lee}{mc1,mc2,mc3}
\icmlauthor{Guillaume Leclerc}{cs1}
\icmlauthor{Joel Dapello}{mc1,mc2,hv}
\icmlauthor{Yug Rao}{pu}
\icmlauthor{Aleksander Madry}{cs1,cs2}
\icmlauthor{James J. DiCarlo}{mc1,mc2,mc3}
\end{icmlauthorlist}
\icmlaffiliation{mc1}{McGovern Institute for Brain Research, MIT}
\icmlaffiliation{mc2}{Department of Brain and Cognitive Sciences, MIT}
\icmlaffiliation{hv}{School of Engineering and Applied Sciences, Harvard University}
\icmlaffiliation{mc3}{Center for Brains, Minds and Machines, MIT}
\icmlaffiliation{cs1}{Computer Science and Artificial Intelligence Laboratory, MIT}
\icmlaffiliation{cs2}{Department of Electrical Engineering and Computer Science, MIT}
\icmlaffiliation{pu}{Purdue University}
\icmlcorrespondingauthor{Chong Guo}{chongguo@mit.edu}
\icmlkeywords{Machine Learning, ICML}
\vskip 0.3in
]
\printAffiliationsAndNotice{}

\begin{abstract}
Visual systems of primates are the gold standard of robust perception. There is thus a general belief that mimicking the neural representations that underlie those systems will yield artificial visual systems that are adversarially robust. In this work, we develop a method for performing adversarial visual attacks directly on primate brain activity. We then leverage this method to demonstrate that the above-mentioned belief might not be well founded. Specifically, we report that the biological neurons that make up visual systems of primates exhibit susceptibility to adversarial perturbations that is comparable in magnitude to existing (robustly trained) artificial neural networks.
\end{abstract}

\section{Introduction}
Deep neural networks (DNN) for computer vision are brittle in that their decisions are sensitive to small image perturbations which are targeted to modifying their outputs (adversarial attacks; \cite{szegedy2014intriguing,goodfellow2015explaining,carlini2017towards}). This is commonly regarded as an area for system improvement since similar brittleness has not been demonstrated in biological vision at comparable image perturbation strengths. Researchers have produced significant progress towards defense algorithms which improved the adversarial robustness of vanilla DNNs in standard computer vision tasks \cite{Qin2019AdversarialRT,madry2018towards}. Adversarial training in particular has been shown to both increase robustness on the original task \cite{croce2020reliable} and to produce internal representations that better support transfer learning \cite{salman2020adversarially,utrera2020adversarially} and image synthesis/manipulations \cite{engstrom2019learning,santurkar2019image,ledig2017photo}. Yet, despite this progress, it is still widely believed that even the best of these networks are less robust than the 'gold standard' -- the primate visual system. Is this assumption correct?  In this study, we checked this assumption using primate neural recordings and ask: are the high level biological neural representations underlying primate object recognition truly more robust than existing artificial neural representations underlying current AI object recognition?
 
\begin{figure*}[t]
\vskip 0.2in
\begin{center}
\centerline{\includegraphics{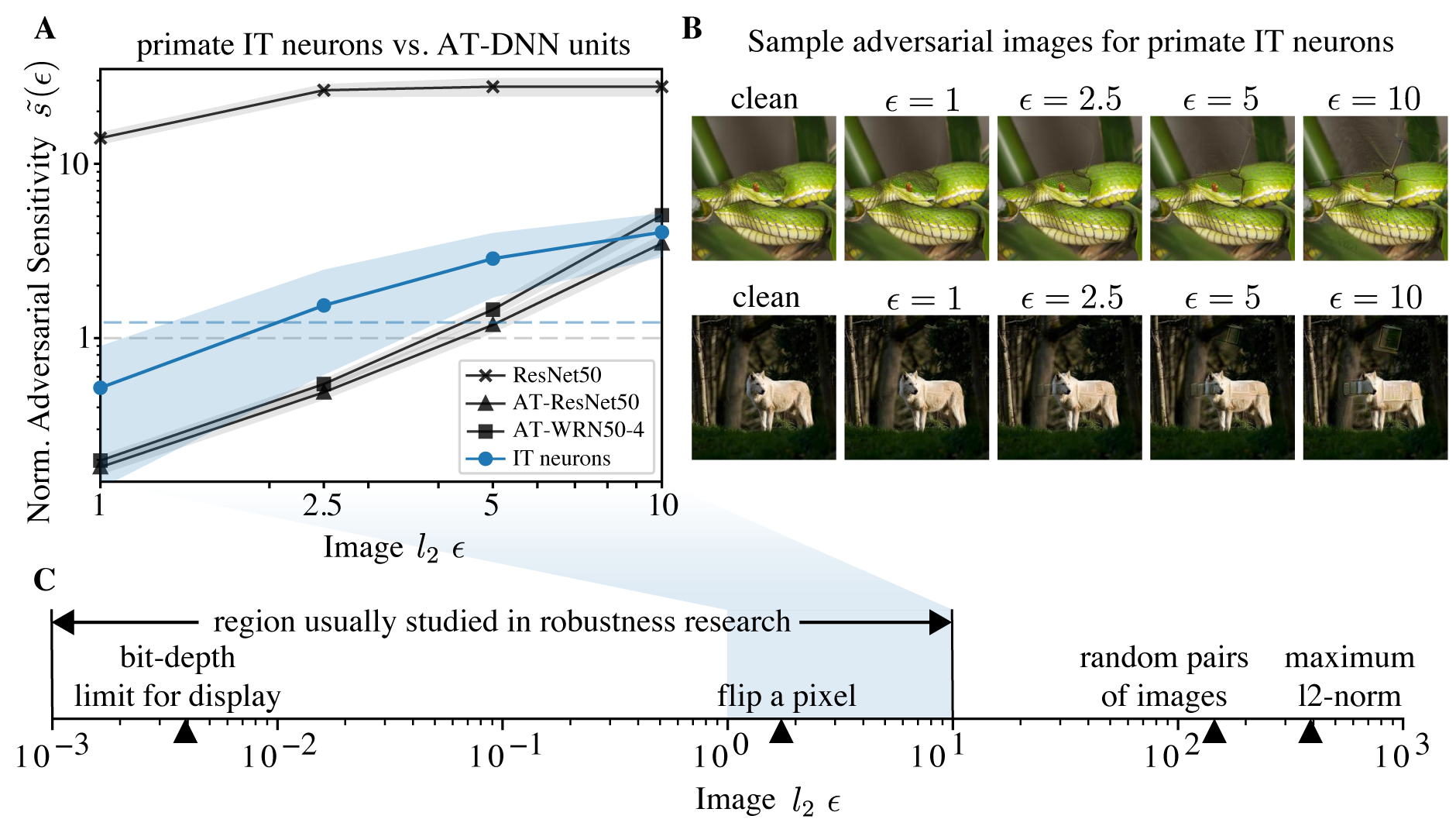}}
\caption{We measured the empirical lower-bound on adversarial sensitivity in biological neurons over a range of $l_2$-norm image perturbations relevant to current robustness research (blue section over the scalebar in \textbf{C}). \textbf{A.} Average normalized adversarial sensitivity $\tilde{s}(\epsilon)$ of IT neural sites (blue, mean$\pm$s.d.) is compared to that of features obtained from standard training on ResNet50, AT-ResNet50 and AT-WRN50-4 with $l_2\epsilon=3$ (black, mean$\pm$s.d.). Standard deviation of either IT neural or DNN unit responses on clean images is shown (grey dashed line). The average absolute difference in IT neural responses evoked by a pair of random images is also shown for scale (blue dashed line). \textbf{B.} Adversarial images for two sample neural recording sites. \textbf{C.} The expanded scale for all feasible $l_2$-norm image perturbations for primate experiments, and notable perturbation sizes (dark triangles) are shown for reference on the horizontal-axis.}
\label{fig1}
\end{center}
\vskip -0.2in
\end{figure*}

It is has not previously been possible to make robustness tests of primate visual neural networks comparable to those in artificial networks because the strongest adversarial attacks approximate the \textit{worse-case} image perturbations by relying on detailed knowledge on each artificial network -- knowledge that is still being developed for the biological system.  Thus when that knowledge is limited, neuroscience experiments must rely on random sampling of image perturbation directions, which -- given the high dimensionality of images -- is unlikely to yield good estimates of adversarial (i.e. worst case) sensitivity within the time constraints of typical primate neuroscience experiments. Indeed, prior neuroscience work measuring the sensitivity of high-level visual representations in primates has focused on a restrictive set of image corruptions (i.e. scrambling) \cite{Rust2010-sj} and quantifying invariance to transformations (position, size, context etc.)\cite{DiCarlo2012-rc,Logothetis1995-du,Ito1995-fj,Tovee1994-fv,Schwartz1983-cx,Sary1993-vo,Ratan_Murty2017-er}. While informative, that work has not directly investigated the adversarial sensitivity of those neurons to small changes in pixel space.   

Here we improved upon recent advances in mechanistic models of primate visual processing \cite{Bashivan2019control,Yamins2014-pa,Dapello2020-rg} to develop an experimental method to efficiently and iteratively measure the lower bound adversarial sensitivity of individual neural sites at the last stage of primate ventral visual processing pathway, the inferior temporal cortex (IT). Primate IT neurons are known to individually encode high-level features and to collectively underlie the perception and behavioral report of visual world latents such as object category and identity \cite{DiCarlo2012-rc,Logothetis1995-du,Tanaka1996-gc,Miyashita1993-sc,Majaj2015-mn, Hong2016-yo}. For these reasons, IT can be loosely considered the functional equivalent of the layer just before the linear soft-max decoder in an artificial neural network. 

For context, prior experimental work on IT neurons focused on much larger image perturbations that were motivated by the hypothesized computational goals of the primate ventral stream (e.g. estimate object category invariant of viewing conditions)\cite{Logothetis1995-du,DiCarlo2012-rc,Rolls2000-at}. Based on such works, we and others in the field thought it unlikely that IT neurons would be sensitive to the much smaller, nearly human imperceptible adversarial perturbations used in current machine learning robustness research (\textbf{Figure 1}). This assumption has resulted in a gap in our knowledge of neural response properties within the local vicinity of any given image. Therefore in this work we attempted to bridge this gap by directly comparing the adversarial sensitivity of individual IT neural sites with individual units in state-of-the-art robust deep neural networks.

\section{Result}
Our primary goal was to measure the sensitivity of the response of individual IT sites to worst-case local pixel perturbations of visual stimuli. For each neural site $i$, we measure its response $r_i(x)$ to clean images $x \sim \mathcal{D}$, where $\mathcal{D}$ is the ImageNet training set \cite{deng2009imagenet}. We define the $i^{\text{th}}$ site's image-specific neuronal adversarial sensitivity as the maximal observed movement that $r_i(x)$ makes under an $l_2$-norm bounded image perturbation $\delta$ which is $\epsilon$ away from the original image $x$: 
\begin{equation}
s_i(x, \epsilon) = \max_{||\delta||<\epsilon} | r_i(x) - r_i(x+\delta)|
\end{equation} 
Marginalizing the image distribution $D$, we define the $i^{\text{th}}$ neural site's absolute adversarial sensitivity as:
\begin{equation}
s_i(\epsilon) = E_{x \sim D}[s_i(x, \epsilon)]
\end{equation}
To be able to compare individual IT neural sites with individual units in artificial neural networks, we use a dimensionless normalized adversarial sensitivity measure $\tilde{s}_i(\epsilon) = s_i(\epsilon)/\sigma_i$, where $\sigma_i$ is the standard deviation of the neural site's response over many different clean images (see \textbf{Method}). Conceptually, this standard deviation is the site's typical dynamic range and because the same response normalization is applied to each artificial neural site, this allows meaningful quantitative comparison between artificial network units (original units: arbitrary scalar response) and biological neural sites (original units: spikes per second). Lastly we note, while the adversarial sensitivity for units in artificial networks are measured on white-box attacks, the attacks for IT neural sites are found using an imperfect model of those sites. Because of this, the estimated adversarial sensitivity for any IT neural site is strictly a lower bound on its white-box adversarial sensitivity (see \textbf{Method}).

We now report the main result comparing the average normalized adversarial sensitivity for primate IT neural sites with the quantitatively comparable measurements of units in the corresponding feature layer of artificial neural networks (\textbf{Figure 1}). For clarity, we reserve the experimental details and the algorithm for empirically lower-bounding the adversarial sensitivity for primate IT neural sites in the method section (\textbf{Figure 4}). 

For reference, we first show the sensitivity of artificial units from the last layer before the linear soft-max decoder in a standard ImageNet pre-trained ResNet50 (black crosses, \textbf{Figure 1A}). The classification performance of this network is highly sensitive to adversarial attacks \cite{robustness}, and as expected, our measure shows that individual units from vanilla ResNet50 are highly susceptible to $l_2$ pixel perturbations: the magnitude of response perturbations are on average over 10-fold larger than each unit's baseline response variations across many clean images (grey dashed line). In comparison with units from ResNet50, we find that individual IT neural sites are approximately 10-fold less sensitive to the same magnitude of image perturbations (blue round markers, n=21, \textbf{Figure  1A}). So far this result is qualitatively consistent with the standard intuition that primate vision is more robust than standard DNNs (above). 

But what about the comparison with adversarially trained (AT) DNNs? Specifically, we tested AT-ResNet50 and AT-WideResNet50-4 after ImageNet adversarial pre-training with $l_2\epsilon=3$. Both networks have improved adversarial robustness on ImageNet and outperform the vanilla ResNet50 on a variety of transfer learning tasks \cite{salman2020adversarially}. Consistent with this, individual units from both AT-ResNet50 and AT-WideResNet50-4 (black triangle and square, \textbf{Figure 1A}) are much less sensitive than units from vanilla ResNet50. Surprisingly however, when we compared to IT neural sites, we discovered that units from both robust DNNs are slightly \emph{less} sensitive to adversarial perturbations than IT sites. This is all the more surprising as our measurements on the biological neurons is an experimental \emph{lower bound} on their true adversarial sensitivity since our attacks are discovered using an approximate model of IT (see \textbf{Method} and \textbf{Figure 4}). On the other hand, the adversarial sensitivity for each unit in the DNNs is measured using white-box PGD-attacks by taking the maximum over 100 randomly initialized attacks with 250 attack steps each (see \textbf{Method}). In other words, further experiments on these IT sites can only lead to the same result -- that current robust network units are already \emph{at least as} robust as these biological network units.
\begin{figure}[ht]
\begin{center}
\centerline{\includegraphics[width=\columnwidth]{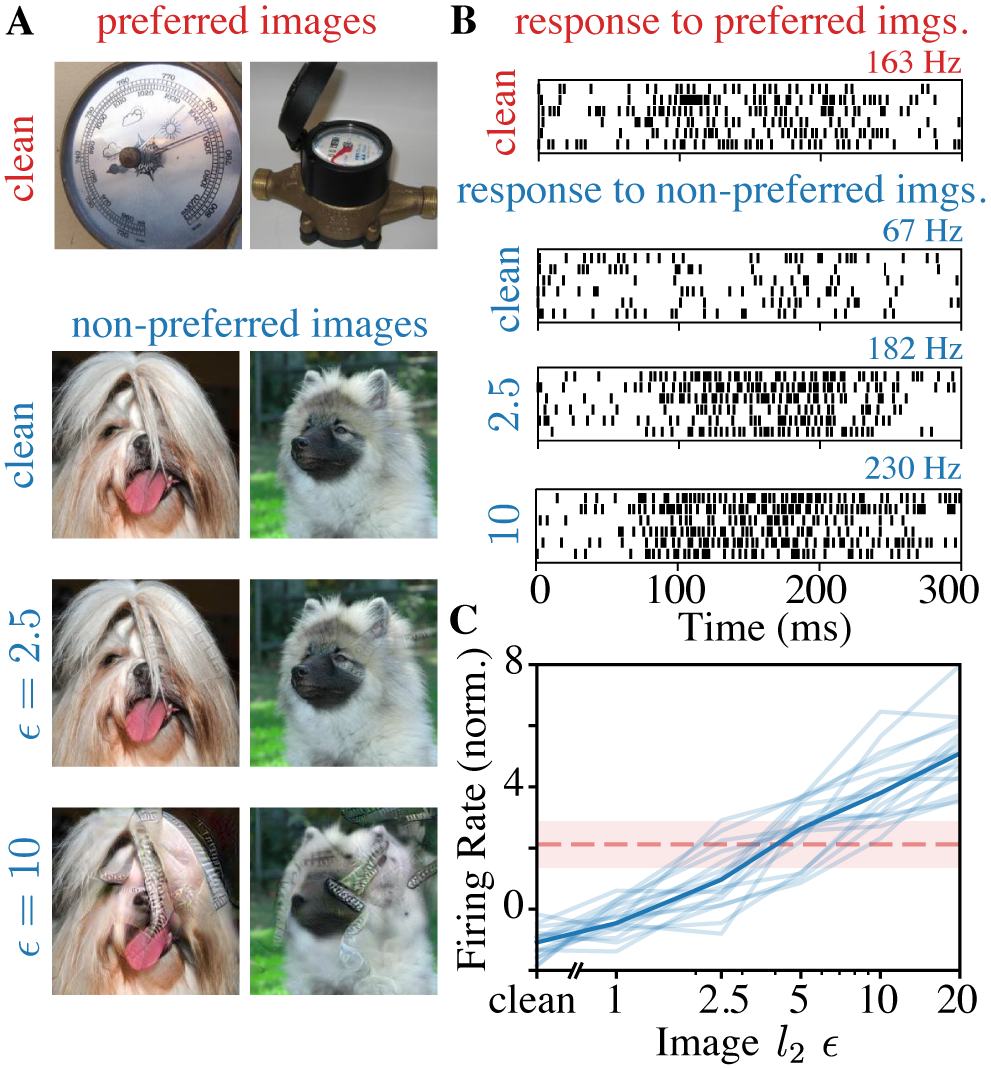}}
\caption{Category preference of individual IT sites can be adversarially attacked. \textbf{A.} Example clean images from preferred and non-preferred ImageNet categories for a representative IT neural site shown on the left, and the adversarially perturbed images at two $\epsilon$ values. \textbf{B.} Raw spike rasters associated with clean preferred images, clean non-preferred images and adversarially perturbed images at $\epsilon=2.5$ and $10$. \textbf{C.} Adversarial perturbations on non-preferred images (in blue) are able to drive firing rates past that of naturally occurring preferred images (red dashed line, $mean\pm s.d.$) at $\epsilon=4.0\pm1.6$ ($2.5$ for the sample site), and by $\epsilon=10$ turn non-preferred images into "super-stimuli" (individual sites in light blue, average in dark blue n=17).}
\label{fig2}
\end{center}
\vskip -0.2in
\end{figure}

The ability to adversarially attack biological neurons at such small perturbations levels can generate some highly counter-intuitive neurophysiological phenomena. Traditionally, IT neurons are known to demonstrate category/object selectivity \cite{Gross1972-sd}. Neurons that respond to images of human faces, hands or specific animate or inanimate objects have been reported throughout IT cortex \cite{Kanwisher1997-jr,Tsao2003-fe,Downing2001-nd,Popivanov2014-oy,Kornblith2013-dk,Bao2020-ii}. 
Although the field no longer regards individual IT neurons as pure object-category detectors, it is still thought that rank-ordered object-category preference is an important single unit property that underlie invariant object recognition \cite{Li2009-no}. Thus we ask the following simple question, how stable is ``category preference'' as a defining functional property of each IT neural site? We used each IT neural site's response over many clean images to identify its most and least preferred ImageNet categories (among 1000 categories), and difference in response between images of preferred vs. non-preferred categories. An example site is shown in (\textbf{Figure 2B}) along with two example images from its most and least preferred categories (\textbf{Figure 2A} top two rows), and the corresponding spike rasters show the clear difference in the density of spiking response following image presentation (preferred clean vs. non-preferred clean \textbf{Figure 2B} top two plots). Now we take arbitrary images from the non-preferred category (i.e. dog images for this example site) and we perform targeted adversarial perturbation on each of those images to change (here, increase) the neural site's response. We find that, by this reference calibration, perturbation slightly over $\epsilon=2.5$, is on average sufficient to turn the site response to any non-preferred image into its same level of response to highly-preferred images  (\textbf{Figure 2A,B} third rows). With a slightly larger perturbation $l2_2\epsilon=10$, we found that we can drive the IT multi-unit site response to $310\pm60$Hz on average (n=17), effectively turning arbitrary non-preferred images into "super-stimuli" for these IT neural sites (\textbf{Figure 2A,B}, last rows). This exceeds the average response of these neurons to their 'most-preferred' image categories ($216\pm60$ Hz). Interestingly, upon visual inspection, the super-stimulus do not usually even conform to the semantic categories from the clean preferred categories. This suggests that single-neuron's 'category/object selectivity' is not a locally stable functional property and highlights the insufficiency of concepts such as 'object detectors' and 'category selectivity' in building an accurate understanding of higher-level visual encoding. 

It is surprising that primate IT neurons, which are approximately six anatomical stages of visual processing deep in the brain, were responsive to perturbations as small as $\epsilon=1$, a barely noticeable change for humans (see \textbf{Appendix 2}) and smaller in magnitude than changing a single pixel from black to white ($\epsilon=1.73$) (\textbf{Figure 1}). Are all IT neurons susceptible to adversarial attacks, or could the average results above be due to just a few strongly modulated neurons? We found that, while the sensitivity level varied, all recorded IT sites were similarly sensitive to adversarial attacks and each has a significantly positive slope between $\epsilon=1$ to $10$ (\textbf{Figure 3A}). We also analyzed the image-specific sensitivity, $s_i(x, \epsilon)$, by recording additional trials (10 repeats) to obtain a cleaner estimate on a subset of 100 images for three neural sites. We illustrate those results by showing sensitivity curves for each of 50 starting images for one typical IT neural site (\textbf{Figure 3B}). Here again, while the sensitivity curves for individual images varied, we found that most had significantly positive slopes measured between $\epsilon=1$ to $10$. Taken together, our results suggest that adversarial images can be readily found on all recorded IT sites and can be found very close to any clean image in the ambient image space (i.e. adversarial samples for biological neurons are dense in the image space similar to that of artificial neural networks \cite{szegedy2014intriguing}).

\begin{figure}[h]
\begin{center}
\centerline{\includegraphics[width=\columnwidth]{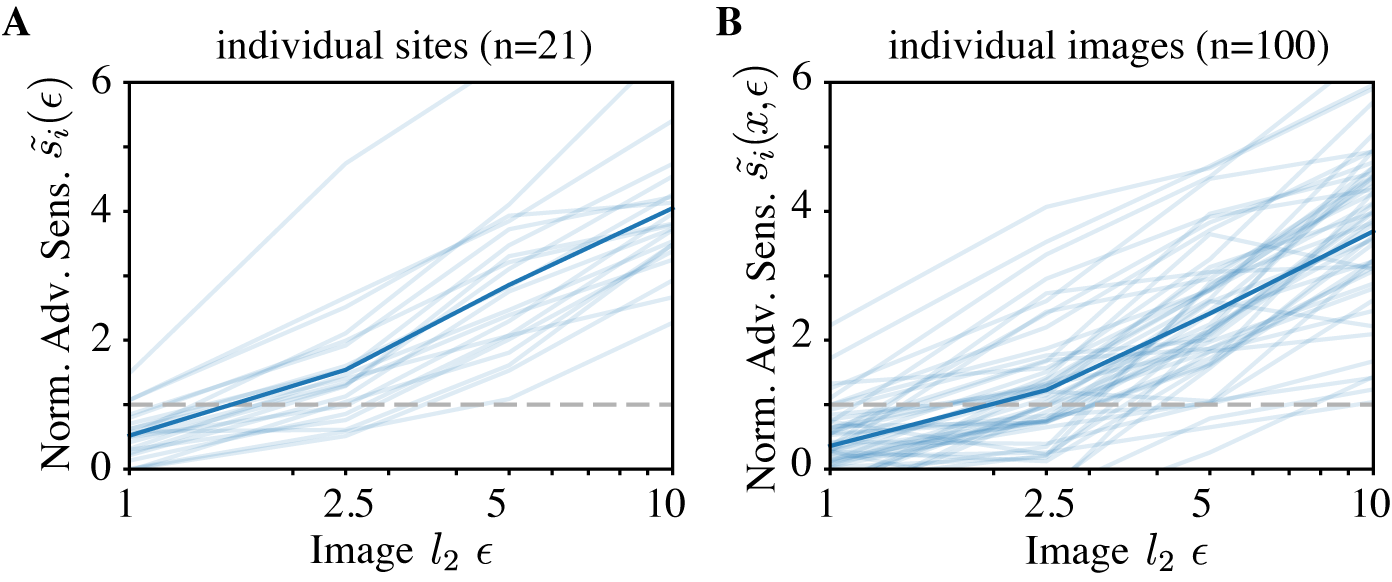}}
\caption{All tested individual IT neural sites are comparably sensitive to these image perturbations, and successfully perturbing images can be found near any starting (clean) image. 
\textbf{A.} Neural site-specific sensitivity curves (light blue) and the average over all sites, reproduced from figure 1 (dark blue). \textbf{B.} Image-specific sensitivity curves for a single example IT site. Each line (light blue) is the site's measured sensitivity to perturbations near a starting clean image (for visibility, 50 of 100 randomly selected starting clean images are shown). The dark blue line shows the average over all images tested for this site (i.e. one of of the light blue lines in A)}
\label{fig3}
\end{center}
\vskip -0.4in
\end{figure}

\section{Method}
\begin{figure*}
\vskip 0.2in
\begin{center}
\centerline{\includegraphics[scale=1.2]{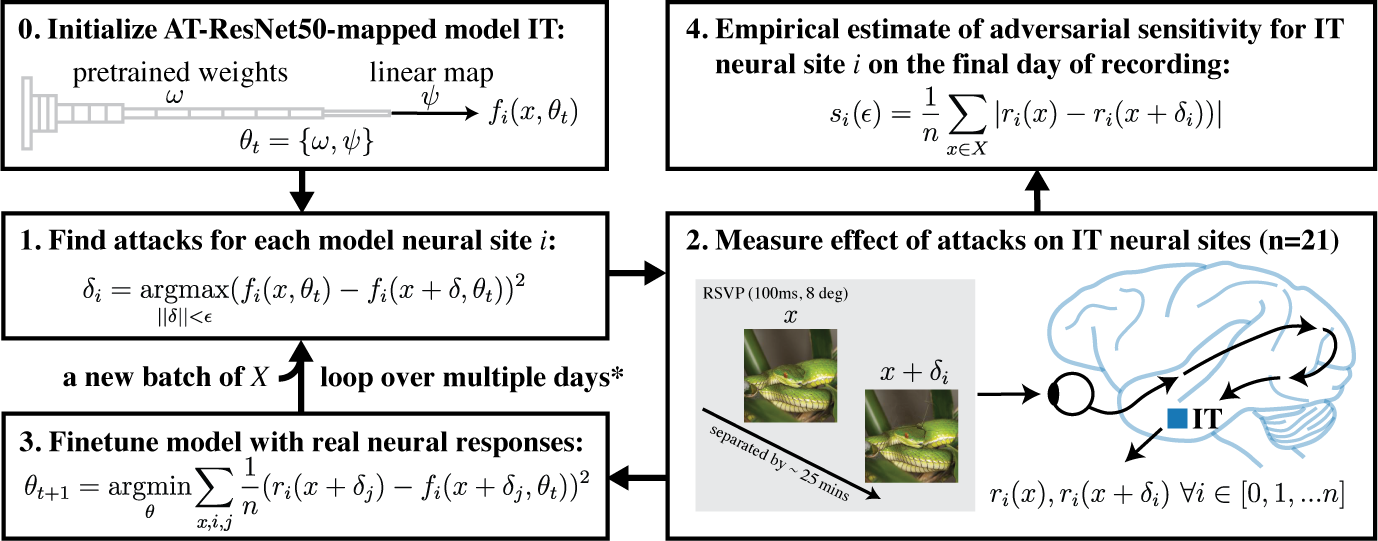}}
\caption{Adversarial training with primate IT neural recordings in the loop allows us to produce a robust functional copy of IT neural sites (box 0-3) and measure the lower-bound on each IT neural site's adversarial sensitivity (box 4)}
\label{fig4method}
\end{center}
\vskip -0.2in
\end{figure*}

\subsection{Measuring adversarial sensitivity of IT neural sites} 
Measuring adversarial sensitivity $s_i(\epsilon)$ for each IT neural site $i$ requires us to maximize the observed $| r_i(x) - r_i(x+\delta_i)|$ by finding better and better neuron-specific perturbation $\delta_i$ on every image $x$ (\textbf{Figure 4}). To do so, we start with perturbations generated from a random baseline model of IT, and iteratively fit the model to observed IT attack responses until convergence. We performed extensive experiments to screen for the best baseline model of the IT sites (see \textbf{Appendix 1} for details on our screening procedure). We found the best baseline model is an adversarially pre-trained ImageNet model AT-ResNet50 ($l2\epsilon=2$) that is linearly-mapped with channel-factorized weights from layer 4.0 to a 21 dimensional output layer to model the IT neural sites (\textbf{Figure 4 box 0}. The model parameters $\theta_t$ at $t=0$ includes the pre-trained model parameters $\omega$ and the linear IT mapping weights $\psi$. Linear IT mapping weights $\psi$ are factorized and initialized as channel weight $w_c \sim N(0, 1)$ and spatial weights $w_s \sim N(0, 1)$ \cite{Klindt2017-pa}.

Using this randomly mapped baseline model, we optimize attack images independently for each model neuron using PGD with random starts, 100 steps and step size=$\epsilon/3$ (\textbf{Figure 4 box 1}). The clean images used on day 0, $X_{t=0}$, consist of 1000 clean images sampled from one of each of the 1000 ImageNet classes from the clean training set. Because for each neuron, the perturbation can either increase or decrease its firing rate, we perform PGD for both loss functions $\text{MSE}(r_i(x+\delta_i),1000)$ or $\text{MSE}(r_i(x+\delta_i),-1000)$ and pick whichever one resulted in the largest predicted magnitude of neural perturbation.

After the attack images are found, we show both clean and attack images to a fixating monkey with two 99-channel Utah arrays (1.5mm, 400 pitch, Iridium Oxide coated electrodes) implanted in anterior and central IT (Blackrock Neurotech, Salt Lake City, UT). The visual stimuli are presented 8 degrees over the visual field for 100ms followed by a 100ms grey mask as in a standard rapid serial visual presentation (RSVP) task. The presentation order for all images (clean and perturbed) are shuffled across the experiment, which lasts for 5 hours on average.  Given this design and the number of images, the average temporal separation between a clean image and its perturbed pair is 25 minutes. Thus image-specific response adaptation is unlikely to explain any aspects of these results. A minimum of two repetitions are shown for each image. For \textbf{Figure 1A}, we report IT sensitivity from the last day of experiment, which sampled 882 unique images per perturbation $\epsilon$ (i.e. 42 images per neural site). For each neural site $i$, we measure the total number of spikes between 70ms-170ms after image presentation for both the clean $r_i(x)$ and perturbed $r_i(x+\delta_i)$ trials (\textbf{Figure 4 box 2}).

\begin{figure}[!h]
\begin{center}
\centerline{\includegraphics[width=\columnwidth]{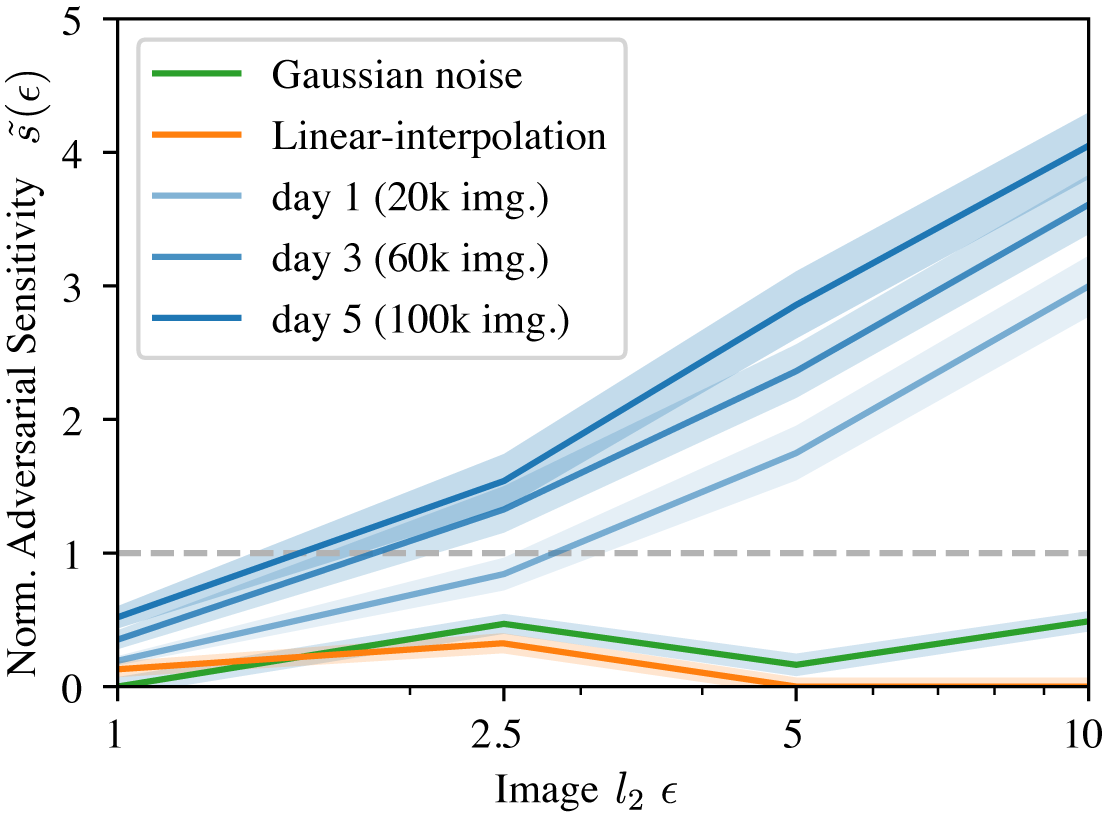}}
\caption{Normalized adversarial sensitivity lower bound improves over a number of days (blue) and is better than model-free perturbations such as Gaussian noise (green) or Linear-interpolation (orange). Sensitivity curve from day 0 where no mapping has occurred is not shown. Model-free measurements are measured as normalized root-mean-squared values which upper bounds the mean absolute differences.}
\label{fig5method}
\end{center}
\vskip -0.35in
\end{figure}

At the end of the first recording session, we use both $(x, \vec{r}(x))$ and $(x+\delta_i,\vec{r}(x+\delta_i))$, $x\in X_{t=0}$, to train $\vec{f}(x,\theta_t)$ end-to-end with gradient descent and update $\theta_{t+1}$ (\textbf{Figure 4 box 3}). After this, a new batch of clean images is queued for the next day ($X_{t+1}$) and we repeat step 1 to 3 over a total of 6 days. As the model of IT $\vec{f}(x)$ improves, the adversarial attack $\delta_i$ solved for each neuron $i$ should become better and better at generating larger perturbations in IT $| r_i(x) - r_i(x+\delta_i)|$ (\textbf{Figure 4 box 4}). Indeed, we tracked the measured neural perturbation magnitude and saw an consistent improvement over days (\textbf{Figure 5}). This suggests that the $l_2\epsilon=1$ attacks from the model IT improved and transferred successfully onto real IT neurons. We note the perturbations achieved with our model is significantly larger than that achieved with a model-free method (Gaussian noise or linear interpolation to another image class), which confirms our intuition outlined in the introduction and explains why the field has systematically underestimated the sensitivity of neurons to local image perturbations. At the smallest perturbation $\epsilon=1$, between the second (day 1: 20k samples for model tuning) and the sixth day of experiment (day 5: 100k samples for model tuning) there was a 2.7 fold increase in realized perturbation size. We report the normalized estimate $\tilde{s}_i(\epsilon)$ from the last day of experiment in all other figures as it is the highest lower bound we could obtain.

Lastly, in the case of biological neurons we use a negatively biased estimator for the average absolute neural response movement to avoid overstating the sensitivity of IT neurons. For $x_j$ where $j=0, 1, ..., M$, out of $M$ number of images: $E_j[|r_i(x_j)-r_i(x_j+\delta_{i,j})|]
= \frac{1}{M}\sum_j \text{sign}(r_i(x_j)-r_i(x_j+\delta_{i,j}))(r_i(x_j)-r_i(x_j+\delta_{i,j})
\geq \frac{1}{M}\sum_j \text{sign}(f_i(x_j, \theta)-f_i(x_j+\delta_{i,j},\theta))(r_i(x_j)-r_i(x_j+\delta_{i,j})$. In the presence of measurement noise, taking the average of the absolute value of response change will result in a positive bias. Therefore, the last equation is used for the estimation of average absolute neural response movement over images for each site. This estimator becomes unbiased if the model of IT site $f_i(x,\theta)$ from the \textit{previous day} predicted all the directions of neural movement correctly. 

\subsection{Measuring adversarial sensitivity of individual artificial neurons} 

In the context of robust machine learning, evaluation of the sensitivity of a network is usually done by quantifying the accuracy of a model on adversarially perturbed images. Here we wish to compare the sensitivity of neurons in IT to sensitivity of units within an deep neural networks. The adversarial sensitivity of a single network unit $i$ is defined as:
\begin{equation}
    \label{eq:nn_sensitivity}
    s_i(x, \epsilon) = \max_{||\delta|| < \epsilon} \left|h_i(x) - h_i(x + \delta)\right|
\end{equation}
where $x$ is a given image, $h_i$ the activation of the $i^{\text{th}}$ network unit of the penultimate layer (i.e. average pooling layer in ResNet50s). We use this layer in particular as this is the layer typically used for transfer learning and insofar contains a high-level representation of the pixel input learned by the network.

Besides taking the best out of 100 independent runs for solving the adversarial images for each unit, we also introduced multiple methods to drastically improve convergence beyond the basic PGD typically used for solving adversarial images:

\begin{itemize}
    \item Optimizing $\max_{||\delta|| < \epsilon} h_i(x) - h_i(x + \delta)$ and $\max_{||\delta|| < \epsilon} h_i(x + \delta) - h_i(x)$ separately significantly reduces the chances to be stuck at saddle point. In most cases the latter one produces better solutions but we always attempt both.
    \item We observed that solving for larger $\epsilon$ converges faster. Therefore, for our evaluation we perform 250 steps of projected gradient descent first with a ball of radius $2\epsilon$ and finally with one of radius $\epsilon$. The relaxation of the first phase dramatically improves exploration of the search space and produces higher quality perturbations.
    \item For each optimization loop of 250 steps, we perform simulated annealing with restarts: we begin with steps of size $\epsilon$ and reduce them by $10\%$ every time no progress is made. This schedule is repeated up to 4 times.
\end{itemize}

\section{Related Works}
Only a small number of studies have explored adversarial phenomena in human and non-human primate \cite{Elsayed2018-fy,Yuan2020-gj,Zhou2019-yd}. None of these directly measured adversarial sensitivity of neural representations in the regime studied by robustness research in the computer vision and machine learning communities ($l_2\epsilon<10$). Yuan et al. attempted to adversarially attack neural and behavioral response in a two-way classification task in the primate using a vanilla ResNet101 \cite{Yuan2020-gj}; however, the range of perturbations used in that study was between $l_2=[21.4, 43]$, all of which are clearly visible to humans and substantially higher than what we explore here.  Another study attempted to use an ensemble of non-robust networks to transfer attack time-limited human behavior on a three-way classification task using $l_{\inf}=32/255$ perturbation, which is equivalent to $l_2 \epsilon=48.7$ \cite{Elsayed2018-fy}. As noted by Tramer et al., $l_2$ budgets greater than $19$ fail to measure semantic similarity in images, and as they demonstrate, generating adversarial attacks to fool human behavior with an $l_2$ budget above $20$ is trivially feasible by pasting in a target object without the aid of any behaviorial or neural models to guided the attack \cite{Tramer2020-vb}.

This is also the first paper where a robust neural network is utilized for biological neural control under a limited budget (\textbf{Figure 2}). All existing works used vanilla networks, DNN or GAN, to synthesize images for single neuron or population control in V4 or IT \cite{Bashivan2019control, Ponce2019-fw, Walker2019-ih}. The benefit of robust neural networks for biological neural control is that no additional image level prior needs to be manually enforced via additional loss functions. Relying on the network itself allows us to leverage additional data to discover the correct image level prior for neural/behavior control. This is possibly why we are able to control neurons with far less image perturbation budget, when all prior work used effectively unlimited budget for image manipulation. 

\section{Discussion}
Here we provided the first experimental demonstration of adversarial sensitivity in biological neurons in a high-order brain area involved in visual object recognition. We find that the representations learned by adversarially trained artificial neural networks have already exceeded that of the corresponding biological neural representation in terms of their individual unit level adversarial robustness. Our results suggest that adversarial examples exist for all IT neural sites and that they are dense in the image space (i.e. nearby any starting image). Moreover, we demonstrated l2-norm perturbations as small as $\epsilon=2.5$ could completely alter the category selectivity of recorded units, casting into doubt a traditional approach that vision scientists have relied on for decades for interpreting and cataloging functional neuronal types in IT. 

 This result confronts us with an apparent paradox: How is it that primate visual perception seems so robust yet its fundamental units of computation are far more sensitive than expected? One distinct possibility is that visual object recognition behavior in primate is actually not robust. This could be potentially explored with an iterative adversarial psychophysics experiment, similar to what we have done here for IT neurons. An alternative explanation is that there is an unknown error-correction mechanism at the population level in IT or in a down-stream area that decodes object identity. These hypotheses can be tested in subsequent experiments. We believe the current line of work could potentially lead to biologically-inspired solutions in ML robustness research, provide fundamental insights into the nature of adversarial phenomena in biological cognition, and perhaps provide new avenues to precisely modulate internal brain states without disrupting daily visual behavior.

\bibliography{madry}
\bibliographystyle{icml2022}

\section{Appendix}
\subsection{Optimizing the adversarial base network for modeling IT}

The training $\epsilon$ and layer for the base model of IT are chosen by three independent criteria: A. global representational similarity to IT as measured by CKA, B. cross-validated linear predictivity for IT responses, and C. how well does perturbations targeted toward a model layer transfers to IT neurons without any explicit mapping between the two systems. A and B are both performed on a separate set of neural recordings consisting of 12k images from the ImageNet training set. The best base models as measured by the three independent criteria are in complete agreement, which is layer 4.0 from $\epsilon=2$ trained adversarial ResNet50. The resolution of detecting the best model appears to be better using the perturbation test of adversarial stimuli. Additionally we make the interesting observation that with AT-ResNet50s, adversarial training generally shifts the layer that is most similar to IT down to the deeper parts of the network with larger and larger training $\epsilon$. 

\begin{figure*}[!ht]
\begin{center}
\centerline{\includegraphics[width=\textwidth]{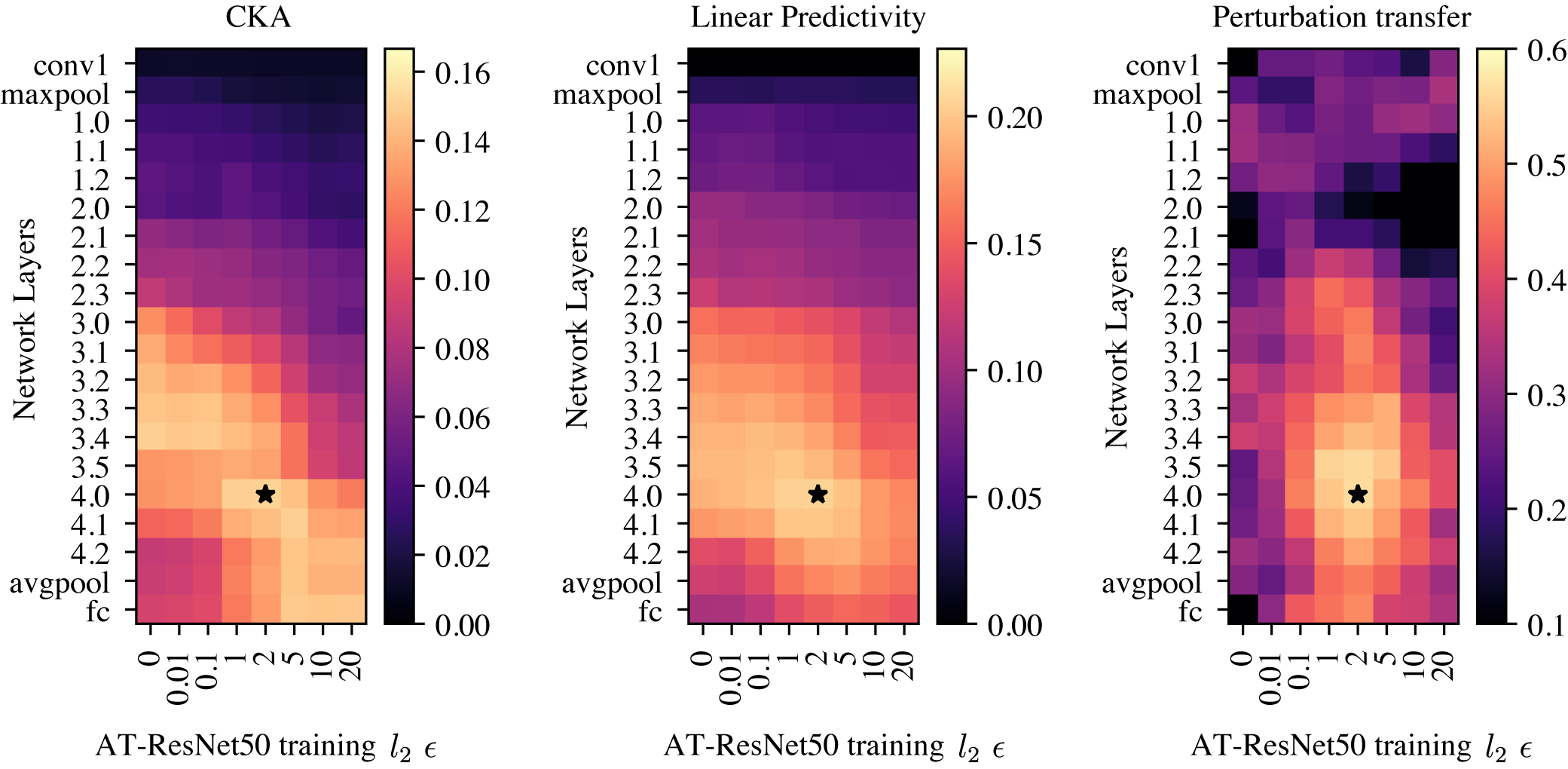}}
\caption{Selecting the layer (vertical-axis) and adversarial training $\epsilon$ (horizontal-axis) for the base mapping model of IT. Black asterisks denote the best layer/training $\epsilon$ combination selected using each of the three metric.}
\label{supp_fig1}
\end{center}
\end{figure*}

\subsection{Perceptual discrimination of adversarial stimuli for IT neural sites} 

In a balanced experimental design where we ask subjects to detect the presence of adversarial perturbations designed for IT neural sites. We find that the accuracy of detecting a perturbation of $\epsilon=1$ to be $51\%$, close to a chance level of $50\%$ of random guessing in our design. Below we describe the detailed experimental protocol for this behavior result.

Briefly, Human subjects (n=50) were recruited on Amazon's Mechanical Turk Platform to conduct our behavioral experiments (n=13,8096 behavioral trials total) following COUHES guidelines. Subjects were free to do as many sessions as they liked (median number of trials per worker = 1,638 trials). 

At the beginning of each session, subjects were given written instructions on how to successfully complete a single trial: first, two images were presented successively (100 msec durations each, with a 100 msec delay in between presentations in which a solid gray background was shown). Then subjects were instructed to report whether the two images were \emph{completely} identical, or not (up to 10 seconds to respond). Images were presented on a neutral gray backgrounds, at approximately 8 degrees of visual field (based on assumptions of typical monitor sizes and viewing distances). Trials in which the subject failed to make a response within 10 seconds were discarded from our analysis. 

On any given session (which consisted of 100 experimental trials), we balanced the number of positive trials and negative trials. Specifically, for any given perturbed image, we included 4 trials: perturbed-clean, clean-perturbed, clean-clean, and perturbed-perturbed. Such a scheme ensured that perturbed images showed up in all positions (first and second frame) and at the same rate as its original version, and that random guessing or choice biases would lead to an average accuracy of $50\%$. Thus, in each session, we were able to obtain an empirical estimate of the true detection rate and false detection rate with respect to 25 perturbed images. 

By collecting data over many such sessions over many images, we were able to obtain subject-averaged estimates of detection rates (corrected for bias) for all of the perturbed images used in this experiment. We were then able to average these estimates by the perturbation $\epsilon$ values associated with each of the perturbed images.

We also included 'catch' trials in which two different base images were presented in sequence (leading to an 'obvious' choice of the two images being different), every 10 trials. By doing so, we could estimate a lower bound on the rate at which subjects were attending to the task (non-attending subjects would have $50\%$ accuracy on catch trials) over the course of the session. On such catch trials, subjects had an average accuracy of $97.4\%$ ($0.95$-CI: [97.1, 97.7]), suggesting the data we collected reflected subjects who understood the instructions and were highly attentive to the task (i.e. were not randomly guessing throughout the session). 

\end{document}